\documentclass[10pt, journal]{IEEEtran}
\label{key}
\usepackage{rtospro}
\usepackage{etoolbox}
\usepackage{booktabs}
\makeatletter

\IEEEoverridecommandlockouts

\date{}
\title{HWE-Bench: Can Language Models Perform Board-level Schematic Designs?}

\newtoggle{fullver}
\toggletrue{fullver}

\newtoggle{revision}
\togglefalse{revision}
\usepackage{xcolor}

\iftoggle{revision}{

}{

}

\begin{document}


\author{Weibo~Qiu, Yinhao~Xiao*, Runyu~Pan*
\IEEEcompsocitemizethanks{\IEEEcompsocthanksitem
Weibo Qiu and Yinhao Xiao are with the Guangdong University of Finance and Economics, China. 
Runyu Pan is with the Shandong University, China. 

(e-mail: \{20191081\}@gdufe.edu.cn, \{rypan\}@sdu.edu.cn).

* Co-corresponding author.
}}


\maketitle

\begin{abstract}
Large Language Models (LLMs) have demonstrated significant potential in various engineering tasks, including software development, digital logic generation, and companion document maintenance.
However, their ability to perform board-level circuit design is understudied, as this task requires a synergized understanding of real-world physics and Integrated Circuit (IC) datasheets, the latter comprising detailed specifications for individual components.
To address this challenge, we propose \hweb, an evaluation framework that benchmarks the ability of LLMs to perform such designs.
It consists of 300 board-level design tasks pulled from open-source and crowdsourcing platforms such as GitHub and OSHWLab, covering 8 application domains, and is complemented with a knowledge base of 2,914 real IC datasheets.
For each task, the LLMs are tasked with generating a schematic from scratch, using the provided circuit functional requirements and a set of component datasheets as input.
The resulting schematic will be checked against a static electrical rules, and then passed to a circuit simulator to verify its dynamic behavior.
Our evaluation show that although current models achieve initial engineering usability and documentation understanding, they lack physical intuition, as the top-performing model achieved an overall pass rate of 8.15\%.
We envision that advancements on \hweb\ will pave the way for the development of practical Electronic Design Automation (EDA) agents, revolutionizing the field of board-level design.

\end
{abstract}



\IEEEpeerreviewmaketitle

\section{Introduction}
\label{s:intro}
Large Language Models (LLMs) have catalyzed a revolution in code generation and autonomous software engineering. 
Pioneering benchmarks, starting from SWE-bench \cite{jimenez2024swebench}, have demonstrated the escalating proficiency of LLMs in navigating complex codebases \cite{siru2025repograph} and executing multi-step coding tasks \cite{zhang2024AutoCodeRover}. 
However, the transition from pure software generation to physical hardware engineering remains a formidable frontier. 
Unlike software development—which primarily relies on abstract logical reasoning and offers high fault tolerance—board-level electronic design is a strict physical discipline. 
It requires the meticulous interpretation of unstructured, multimodal Integrated Circuit (IC) datasheets and strict adherence to electrical and physical exclusivity rules. 
Consequently, current board-level design workflows cannot fully leverage LLM agents; they remain bottlenecked by a reliance on manual human expertise and heuristic practices \cite{Wang2024}.
\begin{figure}[t]
    \centering
    \includegraphics[width=1\linewidth]{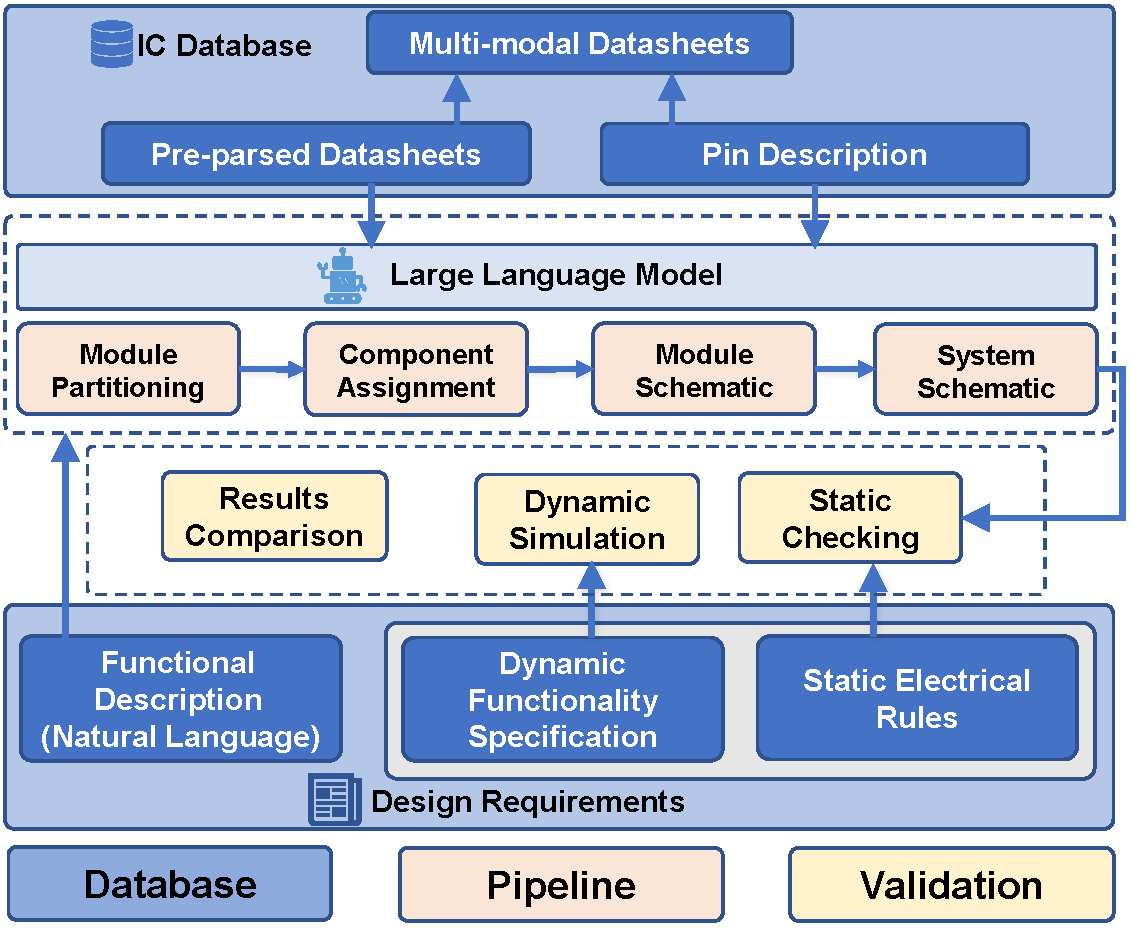}
    \caption{The overall architecture of \hweb. }
    \label{fig:system_architecture}
\end{figure}

To bridge this systemic gap and rigorously evaluate the capabilities of board-level LLM agents, we propose \hweb. 
This comprehensive benchmark framework encompasses 8 prominent application domains, featuring 300 real-world board-level circuit design tasks. 
To authentically replicate the industrial engineering environment, \hweb\ is complemented by a knowledge base of 2,914 real IC datasheets, preprocessed through a dedicated document analysis pipeline \cite{wang2025iclabagent}. 
For each task, the LLM is challenged to generate a schematic from scratch, utilizing only the provided circuit functional requirements and a set of component datasheets. 
Crucially, our pipeline emulates the cognitive workflows of human engineers by utilizing structured prompts with rigorously defined goals. It deliberately directs the model to accomplish key milestones: functional module division, precision component allocation, and local-to-global schematic netlist generation.
To verify the functional correctness of the generated circuits, we abandon traditional text-similarity matching approaches, which are fundamentally ineffective in the zero-fault-tolerant hardware domain. 
Instead, \hweb\ pioneers an execution-driven evaluation: the resulting netlists are first subjected to stringent static electrical rule checking, and subsequently integrated into an industry-standard LTspice simulator to verify dynamic transient behaviors. 
Compared with existing benchmarks, our framework distinguishes itself by:
\begin{inparaenum}[(1)]
\item specifically target board-level designs,
\item enables parsing of multimodal datasheets,
\item performs dynamic simulation in addition to static rule matching, and
\item makes it possible to add new future tasks to the framework with little effort, extending its evaluating capabilities.
\end{inparaenum}

We conducted a comprehensive evaluation of state-of-the-art LLMs (including DeepSeek-v3, Claude Sonnet 4.5, GPT-5.2, Qwen3-Max, Gemini 3.1 Pro, and Grok-4) on our framework. 
The primary contribution of this paper is the establishment of the first execution-driven benchmarking framework tailored for board-level schematic design. 
The paper offers the following contributions:
\begin{itemize}
\item {\bf we collect real-world design tasks}
(\S\ref{s:collect}) from open-source and crowd-sourcing platforms including their task description and IC component datasheets;
\item {\bf we standardize the design tasks}
(\S\ref{s:collect}) to ensure that each of them provide clear task description, machine-checkable goals, and pre-parsed datasheet representations that facilitate performance evaluations;
\item {\bf we propose the test pipeline}
(\S\ref{s:frame}) that emulates the cognitive pathways of human engineers and guides the models to accomplish the design tasks;
\item {\bf we evaluate state-of-the-art language models}
(\S\ref{s:eval}) on our framework, reporting their performance and our findings.
\end{itemize}

\section{Background and Related Work}
\label{s:related}
\subsection{Language model evaluation}
In recent years, the evaluation paradigm for Large Language Models (LLMs) has undergone a profound shift from `static question-answering' to `dynamic execution' \cite{wang2024survey, Zhao2023ExpeLLA}. 
Early benchmarks primarily relied on cross-domain knowledge retrieval or simple text interactions. While subsequent datasets focusing on algorithmic problem-solving have pushed the boundaries of logical generation, these fragmented tests remain confined to theoretical reasoning. 
They struggle to measure a model's true potential in complex engineering contexts and fail to reveal its underlying cognitive mechanisms when navigating multi-step, real-world constraints \cite{Valmeekam2022PlanBenchAE}. 

To bridge this gap, pioneering research has transitioned towards evaluating models within realistic, interactive environments. Frameworks such as WebArena \cite{zhou2024webarena}, AgentBench \cite{liu2023agentbench}, OSWorld \cite{xie2024osworld}, and SWE-bench \cite{jimenez2024swebench} embed LLMs into actual software codebases or operating systems. 
By tasking models with fixing codebase defects or executing system commands, these benchmarks verify capabilities through concrete execution results, aligning LLM evaluation closely with industrial software practices.

However, software environments are fundamentally pure logical abstractions with high fault tolerance, often allowing for iterative trial and error \cite{valmeekam2023planbench}. 
In stark contrast, real board-level electronic circuit design is a typical zero-fault tolerance system governed by strict three-dimensional spatial topologies and physical electrical laws, where a single misconnection can lead to irreversible physical failure \cite{li2026pcbbench}. 
Existing software-centric benchmarks are inherently incapable of measuring a model's ability to cope with such hardcore physical constraints. 
To address this critical void, \hweb\ transcends traditional pure-text and virtual-code testing by integrating expert-level static electrical rule checking with dynamic physical simulations. 
This dual-verification approach establishes a new standard for evaluating and advancing the construction capabilities of LLMs in the tangible physical world.

\subsection{Machine Learning and LLMs in EDA}

Electronic Design Automation (EDA) is a central cornerstone of modern printed circuit board (PCB) design. Due to the natural non-Euclidean graph topology properties of hardware netlists, machine learning models, particularly Graph Neural Networks (GNNs), have become mainstream tools in the EDA field over the past few years \cite{said2023c}. 
They are actively utilized to address board-level spatial optimization and circuit completion tasks \cite{ sath2024aio}. 
For instance, recent studies propose GNN-based node-pair prediction models to automatically add optimizing components, such as pull-up resistors and decoupling capacitors, directly into PCB schematics \cite{plettenberg2025graph}.
Beyond the schematic phase, machine learning has been deeply integrated into physical layout and routing. 
Advanced attention networks have been designed to capture long-range spatial information for thermal-driven PCB routing \cite{Chen2023TRouterTP}, while spatiotemporal graph convolution networks based on PCB layouts are used to predict physical anomalies in solder paste printing \cite{liu2025pcblbs}. 
Furthermore, large-scale network flow datasets like PCBRouteNet have been specifically constructed to accelerate machine learning innovations in automated board routing \cite{ren2025pcbroutenet}. However, GNNs and other heuristic models are essentially discriminative; they rely on existing physical maps for feature extraction at the node or edge level. 
While they excel at spatial mapping, they lack the semantic comprehension required to interpret complex natural language engineering specifications or multimodal datasheets. When faced with cross-modal generative tasks in front-end design, traditional ML models fall short.

Spurred by breakthroughs in logical reasoning, both academia and industry are actively integrating large language models (LLMs) into front-end workflows to realize fully generative board-level hardware design. 
Initial explorations have shown considerable promise, albeit primarily within specific assistive roles. 
For instance, researchers have successfully leveraged models to automate the extraction of multimodal data from complex electronic datasheets, parsing unstructured diagrams into the precise footprint geometries required for PCB component placement \cite{wang2025iclabagent}. 
Alongside this, large language models are being utilized to synthesize PCB test descriptions directly from natural language specifications \cite{li2025llmpcb}, and to guide engineers through intricate board-level routing using few-shot and chain-of-thought prompting techniques \cite{zhang2025pcbr}.

However, while these pioneering studies establish a crucial foundation for AI-assisted board-level EDA, they are fundamentally limited to isolated sub-tasks, assistive text generation, or abstract data parsing. 
In reality, comprehensive board-level design presents a holistic engineering challenge. 
It demands the orchestrated selection of massive heterogeneous components, the meticulous planning of complex electrical topologies, and strict compliance with underlying physical exclusivity rules.
Existing frameworks exhibit a critical capability gap when confronted with these end-to-end generative demands \cite{ma2024ab}. 
To address this systemic void, \hweb\ is proposed as the first comprehensive benchmark dedicated to evaluating an LLM's end-to-end design proficiency. 
It rigorously measures a model's capacity to seamlessly translate unstructured natural language requirements and multimodal datasheet constraints into autonomously generated, globally verified physical-level system netlists.

\subsection{Hardware Generation and Quantitative Benchmarks}

Driven by their remarkable success in software engineering, LLMs are catalyzing a paradigm shift in physical hardware automation, underscoring the immediate need for specialized quantitative benchmarks. 
Broad multimodal frameworks like MMMU \cite{yue2023mmmu} have pioneered the evaluation of expert-level reasoning on electronic schematic diagrams.
In parallel, task-specific benchmarks have been developed to measure localized board-level tasks,notable examples include the PCBRouteNet dataset for learning-based routing \cite{liu2025RouteNet} and the TRouter framework for thermal-aware spatial optimization \cite{chen2023trouter}. 
Additionally, recent studies have formulated rigorous metrics to gauge LLMs' proficiency in interpreting multi-view configurations from datasheets and extracting the complex pinout constraints necessary for automated component selection \cite{Wang2025ALL}.

However, from a systems engineering perspective, generating abstract logic or answering static conceptual queries fails to capture the strict `zero-fault tolerance' demanded by physical hardware realization. Real-world board-level design is a highly heterogeneous discipline. 
It requires the synchronous integration of diverse physical components, such as sensors, operational amplifiers, and power management modules. 
This engineering process necessitates the meticulous handling of continuous analog signals, demanding clock networks, and realistic electrical constraints like voltage limits and power supply topologies \cite{tsou2024auto}. 
Empirical evaluations reveal that even state-of-the-art LLMs face severe limitations in these realistic environments; they frequently violate domain-specific routing constraints in few-shot scenarios \cite{engproc2025aellm} and struggle to generate functionally viable test descriptions for complex PCB assemblies \cite{lidback2025llmpcbtg}. 
Furthermore, while graph neural networks (GNNs) demonstrate utility in localized circuit completion tasks,such as predicting specific optimizing components,they lack the comprehensive semantic depth required for autonomous, system-level schematic synthesis.

The \hweb\ framework fundamentally surmounts the limitations of these localized topological tests and abstract logic evaluations. 
It elevates the scope of LLM hardware evaluation from conceptual text generation to industrial-grade engineering practice, explicitly targeting precise physical component selection and the autonomous generation of complete board-level system netlists. 
By seamlessly integrating expert-level static electrical rule checking with dynamic physical simulations, \hweb\ establishes a rigorous, execution-driven metric for `functional correctness' in the physical world. 
Consequently, it marks a significant milestone toward the realization of fully end-to-end, automated physical hardware design.

\subsection{Execution-Driven Evaluation and Agents}

In the evaluation paradigm of generative artificial intelligence, the field of computer engineering is undergoing a profound transformation from static text analysis to dynamic execution verification. 
While early coding assessments relied on static functional matching, contemporary benchmarks like SWE bench \cite{jimenez2024swebench} require models to resolve real issues within large codebase contexts, determining success through extremely strict unit tests. 
Furthermore, interactive frameworks such as AgentBench \cite{liu2023agentbench} and WebArena \cite{zhou2024webarena} emphasize the multistep execution and self correction abilities of large language models functioning as autonomous agents in embodied environments. 
This shift is motivated by the realization that true autonomy requires environment grounding, where agents must learn and adapt from rigorous environmental feedback rather than static text prompts \cite{wang2024survey}.

In software development, where fault tolerance is inherently high and compile errors can often be iteratively corrected, evaluation driven by execution has become the gold standard \cite{liu2023is}. 
However, in board level hardware engineering, the physical system possesses zero fault tolerance. 
A single misconnected power pin or incorrect analog routing could physically destroy the entire assembly. 
Consequently, similarity evaluations based on traditional static text metrics such as ngram overlaps or abstract syntax tree comparisons are practically meaningless for assessing hardware agents \cite{JIANG2026autocode}. 

\hweb\ inherits and significantly deepens the concept of execution-driven evaluation for the physical domain.
It not only conducts rigorous static pin connection checks using modern graph topology algorithms to verify physical constraints \cite{plettenberg2025graph}, but also seamlessly integrates the generated system-level netlists into industry-standard LTspice simulation environments for dynamic physical level and transient timing tests. 
This dual engine verification establishes a rigorous new standard for evaluating the complex spatial and electrical reasoning capabilities of multimodal agents in the field of board-level hardware engineering.

\section{Task Set Formulation}
\label{s:collect}
\hweb\ is a board-level circuit design benchmark built based on real-world open-source hardware projects.
Its tasks comes from real-world hardware engineering projects in the open-source communities and crowd-sourcing platforms such as GitHub and OSHWLab.
The task is according to given circuit functional requirements and a series of integrated circuit (IC) datasheets, to generate a complete circuit schematic netlist, and pass relevant tests.


\subsection{Repository and Datasheet Collection}
To find high-quality task instances at scale, we use a 2-step pipeline as follows.

\head{Step 1: Repository selection and data scraping.}
We selected board-level circuit design projects from 8 prominent application domains on GitHub and OSHWLab, producing 300 circuit design projects and 2,914 official IC datasheets in total. building a supporting hardware knowledge base.
We focus on prominent application domains as they are representative.
We sort the projects by popularity indicators such as the `Most Likes' in OSHWLab to make sure that they represent the needs on the general market.

As shown in Table~\ref{tab:circuit}, we summarize based on the total number of original pins of all electronic components used in the circuit to distinguish the complexity of the design task. 
This eventually resulted in 300 projects covering 8 areas, each with documentation of functional specifications, bill of materials, and reference designs.
We employ designators to identify the integrated circuits (ICs) and discrete components listed in the bill of materials (BOM), and use a Python script to retrieve the original multimodal datasheets.
A total of 2914 datasheets were obtained, and their distribution is presented in Table~\ref{tab:distribution}.

\begin{table}[htbp]
    \centering
    \caption{Distribution of Circuits across Domains}
    \label{tab:circuit}
    \resizebox{\linewidth}{!}{
    \begin{tabular}{l ccccc}
        \toprule
        \textbf{Domain} & \textbf{[0, 50]} & \textbf{[51, 100]} & \textbf{[101, 150]} & \textbf{[$>$ 150]} & \textbf{Total} \\
        \midrule
        Aerospace   & 1  & 8   & 15  & 2  & \textbf{26} \\
        Educational & 5  & 19  & 12  & 2  & \textbf{38} \\
        Automotive  & 2  & 16  & 6   & 1  & \textbf{25} \\
        Telecom     & 2  & 19  & 13  & 4  & \textbf{38} \\
        IoT         & 1  & 22  & 17  & 4  & \textbf{44} \\
        Consumer    & 4  & 37  & 19  & 8  & \textbf{68} \\
        Healthcare  & 2  & 12  & 8   & 1  & \textbf{23} \\
        Industrial  & 1  & 19  & 13  & 5  & \textbf{38} \\
        \midrule
        \textbf{Total} & \textbf{18} & \textbf{152} & \textbf{103} & \textbf{27} & \textbf{300} \\
        \bottomrule
    \end{tabular}
    }
\end{table}

\begin{table}[htbp]
    \centering
    \caption{Distribution of Datasheets across Categories}
    \label{tab:distribution}
    \begin{tabular}{l cc}
        \toprule
         \textbf{Categories} &  \textbf{Distribution} \\
        \midrule
        Flash & 36.76\% \\
        MCU & 20.86\% \\
        Interface & 12.00\% \\
        Power & 11.32\% \\
        Electromechanical & 11.32\% \\
        Discrete & 4.07\% \\
        Displays & 1.19\% \\
        Other & 1.69\% \\
        \bottomrule
    \end{tabular}
\end{table}

\begin{figure}[t]
    \centering
    \includegraphics[width=1\linewidth]{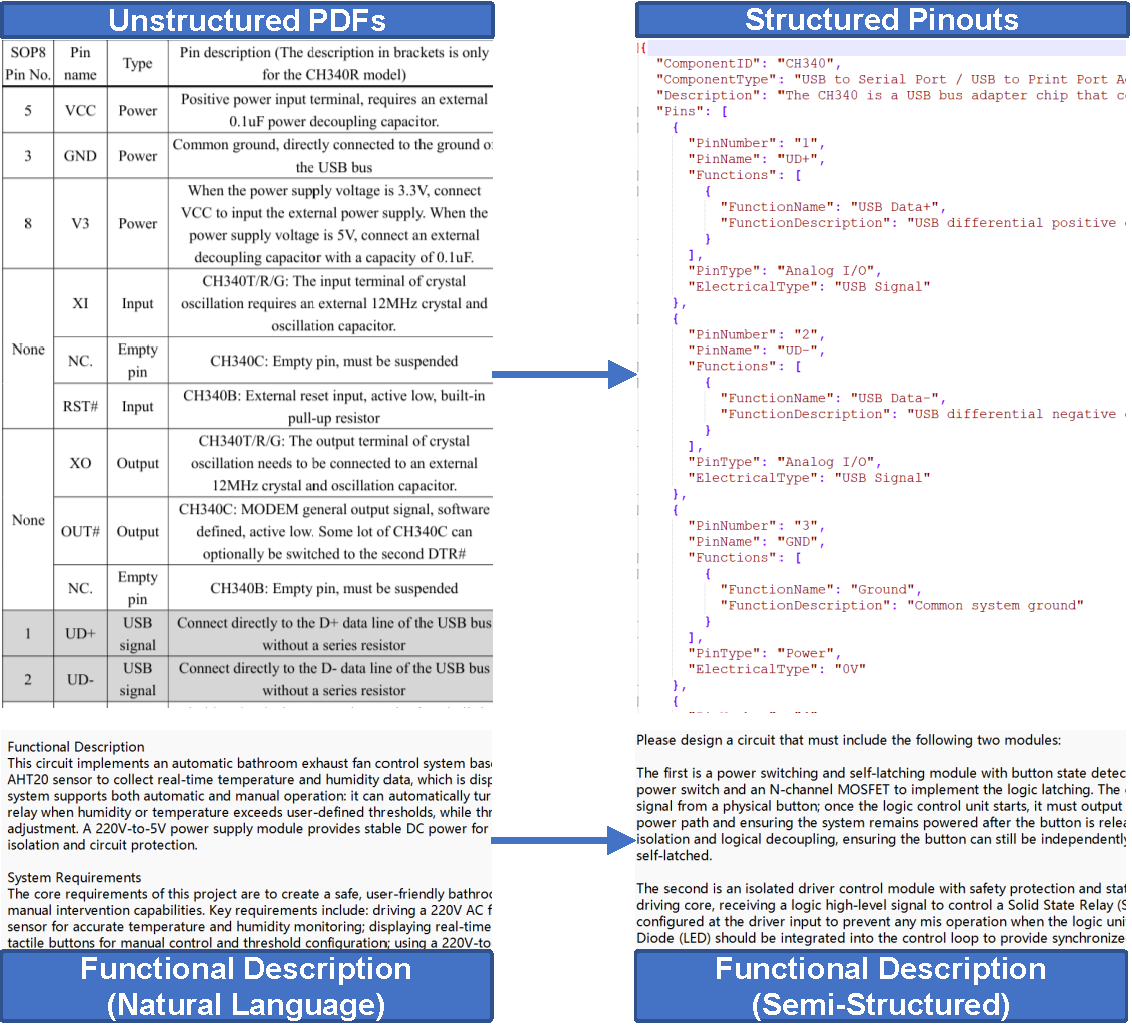}
    \caption{Preprocessing of \hweb. }
    \label{fig:preprocessing}
\end{figure}

\head{Step 2: Repository and datasheet refinement.}
We manually examine the project descriptions from the original repository, and restate the project objectives and constraints in a clear, semi-structured format, eliminating redundancies and ambiguities.
Meanwhile, redundant descriptions from the original open‑source projects were removed, and the precise circuit functional requirements were refined.
As show in Figure~\ref{fig:preprocessing},the pipeline consolidates scattered, unstructured descriptions into coherent project specifications that encapsulate the high-level objectives and outcomes of each project.

For projects lacking sufficient descriptions or providing only limited information, we manually augment the details based on the actual design outputs.
For datasheets, we employed advanced document parsing tools to convert multimodal, unstructured PDF files into Markdown format \cite{wang2024mineru}. 
Subsequently, a guided large language models (LLMs) extracted component descriptions and pin definitions from these Markdown files, and further converted them into structured JSON format to support the LLM’s subsequent inference.

Ultimately, we curated a high-quality structured dataset comprising 300 standardized circuit design tasks and 2,914 cleaned, structured electronic component datasheets.

%

\subsection{Goal and Metric Formulation}
Natural language objectives are typically high-level and ambiguous. 
To establish concrete execution steps that can be faithfully followed by the model, we first manually refine the raw unstructured functional descriptions into semi-structured textual requirements. 
During the evaluation pipeline, the execution is fully automated, yet strictly governed by predefined prompt templates.

\head{Goal formulation.}
To establish machine-verifiable evaluation criteria, we encapsulate specific design goals and physical constraints within the system prompts. 
Given the semi-structured requirements and a library of structured IC datasheets, the LLM is guided through a four-stage constrained generative pipeline. 
First, via prompt directives, the LLMs is tasked with partitioning the target circuit into distinct functional submodules. Each module definition is mandated to capture three key aspects: 
\begin{inparaenum}[(1)]
\item its core functionality, 
\item its interaction with other modules in the signal chain, and 
\item the format of its input/output interfaces.
\end{inparaenum}
The LLMs is prompted sequentially to perform component allocation, module-level local network table generation, and finally system-level global network table integration. 
This prompt-guided decomposition standardizes the evaluation workflow and prevents context overload.

\head{Static checking}
utilizes rule-based label matching to evaluate the structural correctness of the system-level netlist. This step rigorously verifies topological interconnections and ensures protocol logic alignment (e.g., SPI, I²C, UART) across MCUs and peripherals. 

\head{Dynamic simulation}
validates the operational behavior of the circuit. To ensure rigorous evaluation, human experts pre-construct a functionally correct ground-truth SPICE circuit, which is then encapsulated into a Python-driven test script. 
The LLMs-generated netlists are subsequently converted into SPICE format and injected into this Python testbench for automated transient simulation. 
A design is deemed successful if it avoids physical damage and maintains all key node voltages and currents within predefined safety thresholds under specified excitations. 

The core evaluation metric is the aggregate pass rate across these static and dynamic checkpoints, which quantitatively reflects the model's overall performance in board-level circuit design.

\subsection{Features Of \hweb}
Traditional benchmarks usually only involve pure logical reasoning and short function completion. In contrast, \hweb\ delves into the field of electronic engineering, which is strictly constrained by physical laws, giving it the following unique characteristics:

\head{Real hardware engineering tasks.}
The tasks are derived from 300 real-world circuit projects spanning eight major application domains in the open-source community, supported by a comprehensive hardware knowledge base comprising 2,914 authentic datasheets.

\head{High-quality structured data.}
Complex multimodal PDF datasheets are converted into structured JSON representations amenable to automated processing, with redundant information eliminated and precise component specifications extracted.

\head{Decomposed task setting.}
Complex system-level designs are decomposed into functionally distinct submodules, guiding the model to execute the entire design pipeline step-by-step, from functional partitioning to global netlist generation.

\head{Dual-stage verification system.}
A dual-stage verification framework is proposed, integrating static electrical rule checking and dynamic SPICE transient simulation. 
The success criterion is defined as the absence of physical circuit damage and compliance of key node parameters with predefined safety thresholds.

\section{Framework Construction}
\label{s:frame}
After standardizing the task set, we need a system that can guide the model to generate complex designs and objectively evaluate their functional correctness. 
This section will introduce the framework of \hweb, as show in Figure~\ref{fig:system_architecture}, including a generative pipeline that simulates the human cognitive process and a verification engine.

\subsection{Generative Pipeline}
Traditional code generation tasks usually require the model to output the complete result in one go \cite{jimenez2024swebench}. 
However, board-level circuit design involves hundreds or even thousands of pin constraints and cross-domain signal interactions. 
Forcing the model to output the complete netlist in one step often leads to severe local logic gaps and hallucinations \cite{wang2024survey}. 
To address this, \hweb\ constructs a generative pipeline that deeply mimics the cognitive paradigm of experienced human hardware engineers, breaking down the large-scale system engineering into four coherent sub-tasks \cite{wei2022cot}:

\head{Module partitioning.}
The model analyzes the functional requirements of the target circuit and decomposes the entire system into distinct functional submodules, such as power management, communication, MCU, and human-machine interface modules. 
To ensure machine readability, the models are constrained via system prompts to directly output a predefined JSON schema containing the functionality, signal sources, and implementation logic for each module. 
An automated script strips any residual Markdown formatting and serializes the raw output into a structured JSON file.

\head{Component assignment.}
Leveraging the structured datasheet library provided by \hweb, the model selects and allocates appropriate electronic components for each submodule \cite{Wang2025ALL}. 
The LLM is prompted to output a detailed signal chain and a specific JSON array mapping each component to its functional position and pin definitions. 
This output is programmatically extracted and appended to the pipeline's JSON state.

\head{Module schematic.}
Based on the internal signal chain of each submodule, along with the pin definitions of the selected electronic components, the model constructs the corresponding local netlist. 
To mitigate JSON formatting hallucinations when dealing with dense network connections, the model is instructed to output connections using a strict plain-text syntax (e.g., {\small\texttt{netX: Component|Pin|Function}}).
A dedicated regular expression parser then cleans this text, verifies the pin formats, and converts the validated netlist into a structured JSON representation.

\head{System schematic.}
This stage severely tests the model's system-level physical intuition, requiring it to correctly handle the signal logic relationships and power networks between submodules. 
The LLMs integrate the local modules into a global plain-text netlist. 
Finally, an automated parsing engine extracts these unified connections and compiles them into a standardized JSON file, which cleanly encapsulates the complete physical topology ready for metric evaluation.

\subsection{Validation}
In the hardware domain, even an incorrect connection of a single critical wire can lead to physical system damage. 
For netlists generated by the model, rigorous evaluation of their static rule connectivity and behavioral-level functionality is imperative. 
To address this, we have developed an automated verification engine incorporating two evaluation strategies:

\head{Static rule checking.}
Static rules are predefined for each design scheme. 
By abstracting the system-level netlist generated by the LLM into a topology graph based on the component-pin mapping, logical modeling of network connections is achieved \cite{plettenberg2025graph}. 
Using pin property matching and interface protocol consistency verification algorithms, we perform a closed-loop comparison between the generated pin set and the predefined constraint rules.
This comparison allows us to quantitatively evaluate the model’s design completeness with respect to the underlying circuit topology.
This strategy enables accurate identification of logical open-circuit risks and physical connectivity faults within the netlist.

\head{Dynamic spice simulation.}
Since static electrical rule checks fail to capture functional behavior under complex timing conditions, we have developed a dynamic transient simulation verification pipeline for this hardware benchmark.
This process can compile netlists generated by large models into SPICE transient simulation scripts and drive the LTspice engine for rigorous physical-level analysis. 
In order to accurately quantify the logical completeness of the design, we have deployed electrical metric monitoring covering the entire lifecycle during the simulation.
The probe voltage or current at each test point must precisely fall within the preset logic threshold window. 
The task is considered fully physically verified only if the generated circuit topology does not cause any transient overvoltage breakdown, short-circuit overcurrent, or logic deadlock, and all test points meet the qualification criteria.

\section{Evaluation}
\label{s:eval}
In this section we explain how inputs are constructed to run \hweb\ evaluation. 
In addition, wereview the models that we evaluate in this work.

\subsection{Evaluated Models}
We have evaluated a number of the most advanced large language models in the industry on the \hweb\ framework, including Claude Sonnet 4.5, DeepSeek-v3, GPT-5.2, Gemini 3.1 Pro, Qwen3-max, and Grok-4. 
To minimize stochastic variance and ensure deterministic engineering reasoning, we standardize the hyperparameters across all models, setting the sampling temperature to $\text{T} = 0.1$ and the nucleus sampling threshold to $\textit{top\_p} = 0.95$.

\subsection{Result}

We evaluated six  models within the \hweb\ framework. 
The evaluation hierarchy is structured into three progressive dimensions, ranging from fundamental connectivity to system level functionality: static rule checks, dynamic behavior simulations, and global correct design rate.

\begin{table*}[htbp]
    \centering
    \caption{Static Rule Checking Pass Rate}
    \label{tab:static}
    \begin{tabular}{l ccccccc}
        \toprule
        \textbf{Domain} & \textbf{DeepSeek-v3} & \textbf{Claude Sonnet 4.5} & \textbf{Gpt-5.2} & \textbf{Gemini 3.1 Pro} & \textbf{Grok-4} & \textbf{Qwen3-Max} & \textbf{Avg} \\
        \midrule
        Aerospace & 65.30\% & 63.79\% & 63.82\% & 65.47\% & 64.38\% & 61.69\% & 64.08\% \\
        Educational & 71.77\% & 71.66\% & 66.78\% & 75.76\% & 51.83\% & 62.87\% & 66.78\% \\
        Automotive & 82.82\% & 83.25\% & 79.05\% & 79.70\% & 57.97\% & 79.40\% & 77.03\% \\
        Telecom & 69.45\% & 68.95\% & 78.20\% & 58.37\% & 67.34\% & 59.05\% & 66.89\% \\
        IoT & 72.07\% & 82.97\% & 76.04\% & 83.07\% & 63.92\% & 73.61\% & 75.28\% \\
        Consumer & 84.30\% & 78.87\% & 60.78\% & 68.53\% & 66.57\% & 75.59\% & 72.44\% \\
        Healthcare & 72.72\% & 77.25\% & 59.16\% & 68.64\% & 73.85\% & 63.35\% & 69.16\% \\
        Industrial & 79.84\% & 95.11\% & 93.49\% & 86.84\% & 61.82\% & 81.17\% & \textbf{83.05\%} \\
        \midrule
        \textbf{Avg} & 74.78\% & \textbf{77.73\%} & 72.17\% & 73.30\% & 63.46\% & 69.59\% & 71.84\%\\
        \bottomrule
    \end{tabular}
\end{table*}

\begin{table*}[htbp]
    \centering
    \caption{Dynamic Behavior Simulation Pass Rate}
    \label{tab:dynamic}
    \begin{tabular}{l ccccccc}
        \toprule
        \textbf{Domain} & \textbf{DeepSeek-v3} & \textbf{Claude Sonnet 4.5} & \textbf{Gpt-5.2} & \textbf{Gemini 3.1 Pro} & \textbf{Grok-4} & \textbf{Qwen3-Max} & \textbf{Avg} \\
        \midrule
        Aerospace & 66.67\% & 70.00\% & 65.00\% & 56.67\% & 48.33\% & 56.67\% & 60.56\% \\
        Educational & 66.67\% & 61.67\% & 65.00\% & 60.00\% & 56.67\% & 78.33\% & 64.72\% \\
        Automotive & 60.00\% & 51.67\% & 56.67\% & 46.67\% & 60.00\% & 55.00\% & 55.00\% \\
        Teleecom & 40.00\% & 66.67\% & 46.67\% & 48.33\% & 50.00\% & 46.67\% & 49.72\% \\
        IoT & 31.67\% & 56.67\% & 56.67\% & 61.67\% & 31.67\% & 31.67\% & 45.00\% \\
        Consumer & 41.67\% & 66.67\% & 51.67\% & 56.67\% & 50.00\% & 65.00\% & 55.28\% \\
        Healthcare & 50.00\% & 73.33\% & 41.67\% & 73.33\% & 41.67\% & 63.33\% & 57.22\% \\
        Industrial & 76.67\% & 96.67\% & 80.00\% & 76.67\% & 75.00\% & 85.00\% & \textbf{81.67\%} \\
        \midrule
        \textbf{Avg} & 54.17\% & \textbf{67.92\%} & 57.92\% & 60.00\% & 51.67\% & 60.21\% & 58.65\% \\
        \bottomrule
    \end{tabular}
\end{table*}

\begin{table*}[htbp]
    \centering
    \caption{Correct Overall Design Rate}
    \label{tab:correct}
    \begin{tabular}{l ccccccc}
        \toprule
        \textbf{DeepSeek-v3} & \textbf{Claude-Sonnet-4.5} & \textbf{Gpt-5.2} & \textbf{Gemini 3.1 Pro} & \textbf{Grok-4} & \textbf{Qwen3-max} \\
        \midrule
         5.12\% & \textbf{8.15\%} & 6.58\% & 5.16\% & 4.4\% & 5.03\%  \\

        \bottomrule
    \end{tabular}
\end{table*}

\head{Static performance.}
As illustrated in Table \ref{tab:static}, static rule checks primarily assess whether the models successfully implement basic functional pin connections for components. 
At this stage, large language lodels (LLMs) demonstrate preliminary engineering utility, with an overall average pass rate of 71.84\% across all evaluated models.
Claude Sonnet 4.5 delivers the most outstanding performance, achieving an average pass rate of 77.73\%; DeepSeek-v3 follows closely with a pass rate of 74.78\%.
The models achieve a pass rate of 83.05\% in the Industrial domain and 77.03\% in the Automotive domain, significantly outperforming other sectors. 
This is because circuits in these two fields mostly use standardized isolation, driving, and communication buses (such as CAN, RS485), and the related topology paradigms are well-represented in pre-training corpora.
This superior performance may be attributed to the prevalent use of standardized isolation and communication buses in these two fields, which ensures their topological structures have higher exposure within the pre-training corpora.

\head{Dynamic performance.}
When the evaluation criteria shift from static topology to dynamic simulations governed by physical laws , Table \ref{tab:dynamic}, the overall average pass rate drops from 71.84\% in the static phase to 58.65\%. 
This decline indicates that there is significant room for improvement in models regarding circuit design under rigorous physical and electrical constraints.

Under these more stringent physical assessments, Claude Sonnet 4.5 continues to demonstrate the strongest robustness, maintaining its leading position with a pass rate of 67.92\%. 
Qwen3-Max and Gemini 3.1 Pro follow in the second tier, with pass rates of 60.21\% and 60.00\% respectively, showing better adaptability in handling physical timing and transient responses.

\head{Correct overall design performance.}
Table \ref{tab:correct} underscores the harsh reality currently faced by large language models (LLMs) in the hardware EDA domain. 
The global correctness rate mandates that the netlists generated by these models must not only pass both static and dynamic verification but also achieve full system-level functionality without any manual intervention.
The results reveal that even the most capable model, Claude Sonnet 4.5, achieves a global correctness rate of only 8.15\%; GPT-5.2 follows at 6.58\%, while the success rates of all other models hover within an extremely low range of 4\%--5\%.

These data provide intuitive evidence that, although modern LLMs possess formidable capabilities in textual parsing and shallow logical reasoning, they still profoundly lack physical intuition. 
When confronted with board-level circuit designs—where a single minor change can impact the entire system—these models are prone to falling into local optima while neglecting cross-module physical coherence.

\subsection{Experimental Analysis}

We attempt to explore the common and individual characteristics exhibited by large language models in board-level circuit generation tasks from the perspective of their underlying attention mechanisms and pretraining biases. 
We observed a notable pattern: when handling pins with clearly defined functional semantic abbreviations, such as those on microcontrollers (MCU) or USB interfaces (e.g., USBD, USBP, CLK, SDA, TXD, RXD, RST), the connection accuracy of all models is extremely high. 
This is because these conventional names closely align with the strong semantic associations accumulated by the models in pretraining corpora, such as open-source embedded code\cite{br2024code, kocetkov2023the}. 
In contrast, when dealing with pins lacking a single clear semantic, such as “Multiplexed Pins,” or those requiring multiple parallel grounds, like “Redundant Pins,” the models often produce serious omissions or connection hallucinations.  

\hweb, during preprocessing, enforces the transformation of unstructured datasheets into a structured JSON format before delivering them to the large models, greatly mitigating misunderstandings caused by document formatting. 
However, the particularities of hardware system design still challenge the attention limits of the models. 
In real board-level circuits, the total number of pins for global components can easily reach hundreds or even thousands. This extremely large and densely structured key-value pair dataset triggers severe context cognition overload. 
Even the most advanced models experience significant degradation in fine-grained information retrieval capabilities when handling such super-long and highly homogeneous contexts, ultimately leading to frequent loss of context state or parsing errors \cite{liu2024lost}.

Most models score higher on spice simulation pass rate than on static pass rate. 
The underlying reason is that models tend to establish connections that are “functionally and logically equivalent” (usually sufficient for simulation to run successfully). 
However, when faced with rigid static physical engineering specifications (such as soft connections used in certain circuits), models often struggle to follow these implicit physical constraints that lack explicit logic. 
The most important point is that circuit design is not unique \cite{huang2021eda}: even with our greatest brainstorming efforts, we cannot restrict a circuit’s design approach with a single rule, which remains an issue that requires consideration in the future.

\subsection{Case Study}
In this section, through two specific sub-cases, we will demonstrate two key points in this process: 
\begin{inparaenum}[(1)]
\item effectively reducing component disorder within the module,
\item solve the pin multiplexing problem.
\end{inparaenum}

\head{Case 1: Structured signal chain constraints.}
To explore in depth how the structured signal chain constraints preset in \hweb\ rescue the model's physical intuition, we extracted and analyzed the ATmega SolderingStation SMD v2. 
This case involves high-frequency PWM signals, microvolt-level temperature feedback, and 24V high-power MOSFET driving, making it highly representative from an engineering perspective. 

Entity role confusion and label conflicts in baseline tests: Due to the lack of macro-level signal chain constraints, the model experienced severe cross-module entity role conflicts when assigning components. 
In the temperature sensing module, the model assigned the gate label of a power MOSFET to the collector of the driver transistor FMMT619; however, in the subsequent power management module, the model hallucinated and attempted to directly align the same IRLR gate label to the PD3 digital pin of the ATmega328P microcontroller. 
This pin reuse conflict caused by the lack of a global logical perspective led to widespread errors during strict Label Matching rule checks, resulting in a very low label matching degree for the core drive path.  
    
Signal chain constraint intervention and result optimization: To correct this disordered label mapping, \hweb\ forcibly introduced an internal signal chain inference mechanism in prompt engineering \cite{wei2022cot}. 
Under this constraint, the model quickly locked onto the core context of the Heater Power Drive \& Control module and meticulously clarified the physical roles of each entity: the PD3 pin of the ATmega328P is explicitly used only as a 5V PWM signal source, while the IRLR acts as the final 24V high-side switch. 
This signal-flow-based dimensionality reduction strategy completely eliminated the role conflict of the same physical entity across different modules. 

\head{Case 2: Introducing functional group constraints resolves pin multiplexing conflicts.} Pin multiplexing is a core feature of microcontrollers, allowing a single physical pin to take on multiple mutually exclusive functional roles such as GPIO, analog input (ADC), or PWM, depending on the underlying configuration. 
When large language models (LLMs) generate netlists, due to a lack of awareness of physical exclusivity, they often make probabilistic connections based only on functional labels, easily overlooking the physical limitation that a pin can only serve a single role at a time. 
This can result in the same resource being occupied by multiple mutually exclusive networks, leading to severe logical short circuits or hardware protocol misalignment conflicts.

We constructed a design constraint model based on functional groups, partitioning all physical resources of the MCU into dedicated bus groups (e.g., I²C, SPI) and a general GPIO resource pool based on their underlying electrical attributes \cite{Wang2025ALL}. 

This abstraction strategy transcends simple `point-to-point' connectivity. 
Instead, it employs preset interface protocol contracts to mandate that every network connection output by the LLM aligns with specific functional group attributes. By establishing these underlying norms for hardware resource scheduling during the design phase, the system effectively filters low-level logical errors.

The verification system, by introducing a `pin locking' mechanism, simulates the physical allocation of hardware resources in the static audit process, ensuring the uniqueness of each physical entity in the logical chain. 
Whenever a connection in the netlist passes the functional group attribute check, the corresponding physical pin is automatically marked by the system as `locked', and any subsequent rules attempting to reuse that physical resource will trigger a LOCK CONFLICT warning. 
This judgment logic effectively avoids common risks of resource contention and physical short circuits at the static netlist level \cite{plettenberg2025graph}.

\section{Conclusions}
\label{s:conc}
This paper introduces the \hweb\ evaluation framework, which encompasses 300 real-world design tasks spanning 8 application domains and 2,914 integrated circuit datasheets.
To tackle the stringent physical constraints inherent in board-level schematic design, we propose a novel functional group‑based rule verification method for static testing.
Through the adoption of a `pin‑locking' mechanism, we effectively mitigate logical conflicts and resource contention that arise when large language models(LLMs) process MCU pin multiplexing scenarios.
Integrated with dynamic validation via SPICE simulation, our evaluation results identify critical limitations in the physical reasoning capabilities of existing models.

In future work, we will focus on guiding large language models(LLMs) to perform reliable reasoning and generate practically deployable circuit designs.
We will enable  models to master the underlying rules of circuit connectivity and strict physical constraints, while further optimizing intelligent scheduling mechanisms to resolve physical exclusivity conflicts such as pin multiplexing.
In parallel, we will explore diversified evaluation methodologies adaptable to the non‑unique nature of circuit solutions.
Our goal is to bridge the gap between logical reasoning and physical implementation constraints, advancing models toward practical EDA agents with strong physical awareness.


\iftoggle{fullver}{
}{
\input{sections/9_biography}
}

\end{document}